\documentclass[preprint,showpacs,preprintnumbers,amsmath,amssymb,superscriptaddress]{revtex4}

\usepackage{graphicx}% Include figure files
\usepackage{dcolumn}% Align table columns on decimal point
\usepackage{bm}% bold math
\usepackage{color}

\newcommand\Tm{\langle\mathbf{T}\rangle}
\newcommand{\ve}[1][K]{\mathbf{#1}}

\def \equi#1{\mathrel{\mathop{\kern 0pt\sim}\limits_{#1}}} 

\begin{document}

%\title{First-passage statistics: linking between confined and unconfined problems}
%\title{Prefactor in long time persistence for unconfined random walks with stationary increments}
%\title{Beyond scaling  exponents  for long time persistence (of  non-Markovian random walks)}
%\title{Beyond persistence  exponents  for non-Markovian random walks}

\title{Survival probability of stochastic processes beyond persistence  exponents   }

\author{N. Levernier}
 \affiliation{NCCR Chemical Biology,
Departments of Biochemistry and Theoretical Physics, University of Geneva, Geneva, Switzerland}
 
 \author{M. Dolgushev}
 \affiliation{Laboratoire de Physique Th\'eorique de la Mati\`ere Condens\'ee, CNRS/Sorbonne Universit\'e, 
 4 Place Jussieu, 75005 Paris, France}
 
 \author{O. B\'enichou}
\affiliation{Laboratoire de Physique Th\'eorique de la Mati\`ere Condens\'ee, CNRS/Sorbonne Universit\'e, 
 4 Place Jussieu, 75005 Paris, France}
 
 \author{R. Voituriez}
\affiliation{Laboratoire de Physique Th\'eorique de la Mati\`ere Condens\'ee, CNRS/Sorbonne Universit\'e, 
 4 Place Jussieu, 75005 Paris, France}
 \affiliation{Laboratoire Jean Perrin, CNRS/Sorbonne Universit\'e, 
 4 Place Jussieu, 75005 Paris, France}
  
  \author{T. Gu\'erin}
 \affiliation{Laboratoire Ondes et Mati\`ere d'Aquitaine, University of Bordeaux, Unit\'e Mixte de Recherche 5798, CNRS, F-33400 Talence, France }

\begin{abstract} 
For  many stochastic processes,
% compact is more precise but may be a little bit 'brutal' to begin...
 the probability $S(t)$ of not-having reached a target in unbounded space up to time $t$ follows a slow algebraic decay at long times, $S(t)\sim S_0/t^\theta$. This is typically the case of symmetric compact (i.e. recurrent) random walks. 
While the persistence exponent $\theta$ has been studied at length, the prefactor $S_0$, which is quantitatively essential,
 remains poorly characterized,  especially for non-Markovian processes.  
Here we derive explicit expressions for $S_0$  for a compact random walk in unbounded space by establishing an analytic relation with the mean first-passage time of the same random walk in a large confining volume.
%This relation is robust and holds for random walks on fractals and for non-Markovian processes with stationary increments, in one or higher dimensions. 
Our analytical results for $S_0$ are in good agreement with numerical simulations, even for strongly correlated processes such as Fractional Brownian Motion, and thus provide a refined understanding of the statistics of longest first-passage events in unbounded space. 
 \end{abstract}

\maketitle

\noindent {\bf Introduction} 

\noindent In order to determine the time it takes for a random walker to find a target, or the probability that a stochastic signal has not reached a threshold up to time $t$, it is required to analyse the First-Passage Time (FPT) statistics. This has attracted considerable attention from physicists and mathematicians in the last decades \cite{Redner:2001a,Condamin2007,pal2017first,grebenkov2016universal,benichou2010optimal,vaccario2015first} notably because of the relevance of FPT related quantities in contexts as varied as diffusion controlled reactions, finance, search processes, or biophysics \cite{metzler2014first,Berg1985}. 

A single-target first-passage problem is  entirely characterized by the so-called ``survival  probability'' $S(t)$ (the probability that the target has not been reached up to time $t$), or equivalently by the FPT distribution $F(t)=-\partial_t S(t)$.  For a symmetric  random walk  in a confined domain, the mean FPT is in general finite and has been studied at length. This led recently  to   explicit results for broad classes of stochastic processes  \cite{Condamin2005,Condamin2007,Benichou2008,Schuss2007,Guerin2012a}. The opposite case of unconfined random walks is drastically different. In this case, either the walker has a finite probability of never finding the target (non-compact random walks), or it  reaches it with probability one (compact random walk) and the survival probability decays algebraically with time, $S(t)\sim S_0/t^\theta$, with $\theta$ the persistence exponent that does not depend on the initial distance to the target. In this case the mean FPT is often infinite so that the  relevant observable to quantify FPT statistics is the long time  algebraic decay of the probability $S(t)$ that the target %/threshold 
has not been reached up to $t$. This, additional to the fact that $\theta$ can be non-trivial for non-Markovian random walks, has triggered a considerable amount of work to characterize  the persistence exponent $\theta$ in a wide number of models of non-equilibrium statistical mechanics. Indeed,  $S(t)$ is an essential observable to quantify the kinetics of transport controlled reactions and the dynamics of coarsening in phase transitions in general~\cite{Majumdar1999,ReviewBray}.

However, if one aims to evaluate the time $t$ to wait for observing a first-passage event with a given likelihood, or to determine the dependence of the survival probability on the initial distance to the target,  one needs to know the prefactor $S_0$, which turns out to be much less characterized than the persistence exponent $\theta$. Even for Markovian random walks this problem is not trivial \cite{Aurzada2015}, as exemplified by recent studies for one dimensional Levy flights \cite{Majumdar2017}, while only scaling relations for $S_0$ (with the initial distance to  the target) are known~\cite{meroz2011distribution} in fractal domains.
%Even for Markovian (memoryless) processes, only scaling relations (with the initial distance to  the target) are known~\cite{meroz2011distribution}. \tb{One notable exception is given by the case of first-crossing for one-dimensional Levy Flights, which has been investigated very recently .} 
However, if the dynamics of the random walker  results from  interactions with other degrees of freedom, the process becomes non-Markovian and the determination of $S_0$ becomes much more involved~\cite{VanKampen1992}. In this case, the only explicit results are derived  from  perturbation expansion around Markovian processes~\cite{delorme2015maximum,delorme2016perturbative}, or have been obtained for particular processes such as  ``run and tumble'' motion (driven by telegraphic noise \cite{Masoliver1986}) or the random acceleration process \cite{Burkhardt1993}. For long range correlated processes, such as fractional Brownian Motion, the existence of $S_0$ is not even established rigourously \cite{Aurzada2011,Aurzada2015}, and it has been found that straightforward adaptation of Markovian methods can lead to order-of-magnitude overestimations of $S_0$  and even to erroneous scalings~\cite{Sanders2012}.  
 
In this article, we rely on a non-perturbative strategy to determine $S_0$, which is of crucial interest to quantify the statistics of long FPT events. Our main result is a relation between the prefactor $S_0$ in the long time survival probability in free space and the mean FPT for the same process in a large confining volume. %In particular, these quantities share the same dependency with geometrical parameters such as the initial distance to the target. 
Our formula thus shows how to make use of  the wealth of explicit results obtained recently on first-passage properties in confinement~\cite{Condamin2005,Condamin2007,Benichou2008,Schuss2007,Guerin2012a,guerin2016mean} to determine
 the decay of the free-space survival probability.
This formula is shown to be robust and holds for Markovian or non-Markovian processes with stationary increments, that are scale invariant at long times with diverging moments of the position, 
in one or higher spatial dimensions, and also for processes displaying transient aging (i.~e. processes with finite memory time, whose initial state is not stationary, see below). 
This theory is confirmed by simulations for a variety of  stochastic processes, including highly correlated ones such as Fractional Brownian Motion.

 \vspace{1cm}
\noindent {\bf   Results}

\noindent {\bf   Markovian case.}  We  consider a symmetric  random walker of position $\ve[r](t)$ moving on an infinite discrete lattice (potentially fractal) of dimension $d_f$ [see Fig.~1(a) for the continuous space counterpart] in continuous time $t$, in absence of external field. The initial position is $\ve[r]_0$. We assume that the increments are stationary (no aging), which means in particular that $\sigma(t,\tau)\equiv\langle \vert \ve[r](t+\tau)-\ve[r](t)\vert^2\rangle$ is independent of the elapsed time $t$. Note that in the case of fractal spaces, we use the standard  ``chemical'' distance defined as the minimal number of steps to link two points on the lattice.  
We define the walk-dimension $d_w$ such that $\sigma(t,\tau)\propto \tau^{2/d_w}$ for $\tau\to\infty$.   Note that (i) this scale invariance is assumed only at long times, and that (ii) it implies that all even moments of the position diverge with time. We assume $d_w>d_f$ so that the process is compact \cite{DEGENNES1982,benAvraham2000} (and eventually reaches any point with probability one). We also introduce the Hurst exponent $H=1/d_w$.   

We first consider the case of Markovian (memoryless) random walks. One can then define a propagator $p(\ve[r],t\vert \ve[r]_0)$, which represents the probability to observe the walker at site $\ve[r]$ at time $t$ given that it started at $\ve[r]_0$ at initial time. Note that $p$ is defined in absence of target. We now add an absorbing target at site $\ve[r]=0$ (different from $\ve[r]_0$). We start our analysis with the standard renewal equation \cite{VanKampen1992,Redner:2001a,hughes1995random}:
\begin{equation}
p(\ve[0],t\vert \ve[r]_0)=\int_{0}^{t} d\tau F(\tau; \ve[r]_0) p(\ve[0],t-\tau | \ve[0]),
\label{renbasis}
\end{equation}
which relates the propagator $p$ to the FPT distribution $F$ that depends on $\ve[r]_0$. This equation is obtained by partitioning over the FPT to the target, and can be rewritten in Laplace space as
\begin{equation}
\tilde{p}(\ve[0],s\vert \ve[r]_0)=\tilde{F}(s;\ve[r]_0)\tilde{p}(\ve[0],s | \ve[0]),
\label{renLap}
\end{equation}
where $\tilde{F}(s)=\int_0^\infty dt F(t)e^{-st}$ stands for the Laplace transform of $F(t)$. 
Here, we only focus on the long-time behavior of $F(t)$, that can be obtained by expanding Eq.~\eqref{renLap} for small $s$. Scale invariance at long times implies~\cite{benAvraham2000} that  for any site $\ve[r]$
%Since an initially localized probability density function (pdf) will spread on a region of size $t^{1/d_w}$ at time $t$, we have  
\begin{align}
p(\ve[0],t\vert \ve[r]) \underset{t\to\infty}{\sim} K/t^{d_f/d_w}  \label{LongTimePropag},
\end{align}
where the notation   ``$\sim$'' represents asymptotic equivalence, and $K$ is a positive %position independent 
coefficient. Note that $K$ is known to be position independent and  is well characterized (at least numerically) for a large class of stochastic processes, including diffusion in a wide class of fractals    \cite{grabner1997functional,kron2004asymptotics,weber2010random,meroz2011distribution}. 
We find that the small-$s$ behavior of the propagator is
\begin{align}
\frac{K \,\Gamma(1-\frac{d_f}{d_w})}{s^{1-d_f/d_w}} -\tilde{p}(\ve[0],s\vert \ve[r])
\underset{s\to0}{\sim}
\int_0^{\infty} dt \left[\frac{K}{t^{d_f/d_w}} -p(\ve[0] ,t\vert \ve[r])\right],\label{05842}
\end{align}
where $\Gamma(\cdot)$ is the Gamma function. Eqs.~(\ref{renLap}) and (\ref{05842}) (written for $\ve[r]=\ve[0]$ and $\ve[r]=\ve[r]_0$) lead to
\begin{equation}
1-\tilde{F}(s;\ve[r]_0)\underset{s\to0}{\sim}\int_0^{\infty} dt \left[ p(\ve[0],t|\ve[0] ) -p(\ve[0],t\vert \ve[r]_0) \right]  \frac{s^{1-d_f/d_w}}{K\,\Gamma\left(1-\frac{d_f}{d_w}\right)}.
\end{equation}
Taking the inverse Laplace transform (and using $F(t)=-\dot{S}$) leads to $S(t)\sim S_0/t^{\theta}$ with $\theta=1-d_f/d_w$ (as found in \cite{meroz2011distribution}), and to
\begin{align}
&S_0=\frac{\sin(\pi d_f/ d_w)}{K \pi } \int_0^{\infty} dt \left[ p(\ve[0],t|\ve[0]) -p(\ve[0],t\vert \ve[r]_0) \right].
\label{prefMark}
\end{align}
This expression is exact and characterizes the decay  of the survival probability of unconfined scale invariant Markovian random walks.  

We now consider the target search problem for the same random walk, with the only difference that it takes place in a confining volume $V$ (that is equal to the number of sites $N$ in our discrete formulation) [see Fig.~1(b)]. For this problem, the mean FPT $\Tm$ is in general finite and it is  known that it scales linearly with the volume and reads  \cite{Condamin2005,Condamin2007}
\begin{equation}
\frac{\Tm}{V} \underset{V\to\infty}\sim \int_0^{\infty} dt \left[ p(\ve[0],t|\ve[0]) -p(\ve[0],t\vert \ve[r]_0) \right].
\label{TMmark}
\end{equation}
We recognize in the above expression the time integral of propagators appearing in Eq.~\eqref{prefMark}, leading to
\begin{equation}
S_0=\frac{\sin(\pi d_f / d_w)  }{\pi\ K  }\overline{T}, \hspace{0.5cm} \text{with} \hspace{0.5cm} \overline{T}=\underset{V \to \infty}{\lim} \Tm/V.
\label{prefMarkT}
\end{equation}
Hence, for compact Markovian random walks, we have identified a proportionality relation between the prefactor $S_0$ that characterizes the long time  survival probability in free space and the rescaled mean FPT to the target in unconfined space. The proportionality coefficient involves the walk-dimension $d_w$  and the coefficient $K$ which characterizes the long time decay of the propagator [see Eq.~\eqref{LongTimePropag}]. Formula (\ref{prefMarkT}) is the key result of this paper. As we proceed to show, it is very robust and is not limited to Markovian walks.

As a first application, consider the case of scale invariant Markovian random walks (such as diffusion on fractals), for which it was shown \cite{benichou2008zero} that $\overline{T}\simeq r_0^{d_w-d_f}$, where the mean waiting time on a given site is taken as unity, and $r_0$ is the initial source-target (chemical) distance. Inserting this formula into Eq.~(\ref{prefMarkT}) thus leads to
\begin{align}
S_0\simeq\frac{\sin(\pi d_f/ d_w) r_0^{d_w-d_f}}{\pi\ K  } \label{ZeroConstantS0}.
\end{align}
In this case, we thus recover the scaling result of Ref.~\cite{meroz2011distribution} but in addition obtain the value of the prefactor. 
We have checked this relation for the Sierpinski gasket: simulation results are shown on Fig.~2(a). 
The long time persistence is perfectly described by our formula without any fitting parameter for different source-target distances, confirming the validity of our approach (see SI  for other examples). 

%\tr{I like the following example of $\alpha$ processes but strictly speaking it is for continuous space, c'est grave ? Je pense que non mais ça vaut peut-etre une phrase. C'est d'ailleurs etonnant qu'on ait considere que les reseaux pour la cas markovien. A voir avec Olivier et Raphael ?}
As a second application, we can consider the 1-dimensional L{\'e}vy stable process of index $\alpha$, which is defined as the only Markovian process whose jump distribution is given by $p(\Delta x,t)=1/(2\pi)\int_{-\infty}^{\infty} e^{i\Delta x.k-t|k|^\alpha} dk$. This process, defined in continuous space,  is the continuous time limit of the L{\'e}vy Flight with same index $\alpha$. Its walk dimension is  $d_w=\alpha$ and it is compact for  $\alpha>1$, so that the first passage to a point target is well defined (note that we consider here the first arrival at the target, and not the first crossing event \cite{al:2003,Tejedor2011}). For such a process, the prefactor $S_0$ for an unconfined random walk starting at a distance $r_0$ from the target has been shown to be $S_0=\alpha\sin(\pi\alpha/2)\sin(\pi/\alpha)\Gamma(2-\alpha)r_0^{\alpha-1}/(\pi \Gamma(1/\alpha)(\alpha-1))$ \cite{Blumenthal1961}. By computing the rescaled MFPT in confinement with Eq. \eqref{TMmark}, one can check that the relation \eqref{prefMarkT}, which can be readily generalised to  continuous space,  is still verified for this process. 

\vspace{1cm}

\noindent {\bf Extension to non-Markovian processes.} We now relax the Markov property and generalize our theory to the case of non-Markovian processes, i.~e. displaying memory. 
%This problem is notoriously difficult, as noted for instance in \cite{Sanders2012}, where it was shown  that a straightforward use of pseudo-Markovian arguments [such as the image method, or the use of the renewal equation (\ref{renbasis}) with ``effective'' propagators] leads to important numerical errors and can even provide wrong persistence exponents. 
In the following, we argue that the relation (\ref{prefMarkT})  yields much more accurate results for $S_0$ than Markov approximations; it is exact for processes with finite memory time, and is very precise (even though not exact) for strongly correlated processes such as the Fractional Brownian Motion. As the mean FPT in confinement has recently been characterized for non-Markovian Gaussian processes~\cite{guerin2016mean}, this equation  (\ref{prefMarkT})  provides a means to estimate $S(t)$ at long times, beyond persistence exponents, for a wide class of random walks with memory. 
 
 For simplicity, we consider 1-dimensional processes and we switch to continuous space description. The  stochastic trajectories $x(t)$ are assumed to be continuous but non-smooth (the method in fact also applies to compact and not continuous processes, such as $1d$ Levy stable processes  of index $\alpha>1$ as discussed above), mathematically meaning that $\langle \dot{x}^2\rangle=\infty$ and physically corresponding to very rough trajectories, similar to those of Brownian motion. We assume that the increments of the walk are stationary (meaning that there is no aging, even transient \footnote{In particular the case of continuous time random walks (CTRWs) is not directly covered by our analysis;  persistence exponents and prefactors for CTRWs can be obtained from the subordination principle}). This hypothesis is known to have two  consequences: (i) the persistence exponent for the unconfined problem is exactly given by $\theta=1-1/d_w$  \cite{levernier2018universal,Molchan1999,ReviewBray,Krug1997,Aurzada2011,Aurzada2015}; (ii) the mean FPT for the confined problem varies linearly with the confinement volume $V$, so that $\overline{T}$ is finite and has been identified as \cite{guerin2016mean}:
\begin{equation}
\overline{T}=\int_0^\infty dt \,[p_\pi(0,t)-p(0,t)].
\label{Tm} 
\end{equation}
Here, $p(0,t)$ is the probability density of $x=0$ at a time $t$ after the initial state (where $x(0)=r_0$), and $p_\pi(0,t)$ denotes the probability density of finding the walker on the target  at a time $t$ after the first-passage:
\begin{align}
p_\pi(0,t)= \int_0^\infty d\tau \, p(0,t+\tau |  \mathrm{FPT}=\tau) F(\tau;r_0),\label{pPi}  
\end{align}
where $p(0,t+\tau |  \mathrm{FPT}=\tau)$ is the probability density of $x=0$ at time $t+\tau$, given that the FPT is $\tau$. 

The starting point to relate $\overline{T}$ to $S_0$ consists in writing the generalization of Eq.~(\ref{renbasis}) to non-smooth non-Markovian processes:
\begin{align}
p(0,t)=\int_0^t d\tau F(\tau;r_0) p(0,t  | \mathrm{FPT}=\tau).\label{renNM}
\end{align}
To proceed further, we insert into Eq.~\eqref{Tm} the expressions (\ref{pPi}) and (\ref{renNM}) of $p_\pi(0,t)$ and $p(0,t)$:
\begin{align}
\overline{T}=\int_0^{\infty} dt \,\Big[ & \, \int_0^{\infty} d\tau \, p(0,t+\tau \, | \, \text{FPT}=\tau)\, F(\tau;r_0)  \nonumber \\
&-\int_0^{t} d\tau \, p(0,t \, | \, \text{FPT}=\tau)\, F(\tau;r_0) \Big] .
\end{align}
To avoid  diverging integrals in the change of variables $t=u+\tau$, we replace $\int_0^\infty dt(...)$ by $\underset{A \to \infty}{\lim}\;\; \int_0^{A} dt (...)$, so that 
\begin{equation}
\overline{T}=\underset{A \to \infty}{\lim} \int_0^{A} dt \,   \int_{A-t}^{\infty} d\tau F(\tau;r_0)  p(0,t+\tau  |\text{ FPT}=\tau) .
\label{eqA1}
\end{equation}
Setting $t =uA$ and $\tau=vA$, we note that when  $A\to \infty$, only the large time behavior are involved in these integrals, where one can use the asymptotics $F(Av;r_0)\sim S_0 \theta /(Av)^{1+\theta}$ and
\begin{align}
p(0,(u+v)A| \text{FPT}=vA)\underset{A\to\infty}\sim \frac{K G(u/v)}{(Au)^{1/d_w}}\label{DefG},
\end{align}
which is a form imposed by dimensional analysis. As previously, $K$ is the constant which characterizes the long time behavior of the one point probability distribution function [i.e. $p(x,t)\underset{t\to\infty}\sim K/t^{1/d_w}$],
$G$ is a scaling function, with $G(\infty)=1$, that does not depend on the geometrical parameters of the problem.   
Inserting these asymptotic behaviors into Eq.~\eqref{eqA1}, we get:
\begin{equation}
\overline{T}=\underset{A \to \infty}{\lim} A^{1-\theta-1/d_w} \,K \, \int_0^{1} du \,   \int_{1-u}^{\infty} dv \,\frac{\theta \;S_0}{v^{\theta+1}} \frac{G(u/v)}{u^{1/d_w}}
\label{eqThetaetS}.
\end{equation}
The fact that the above integral exists and is finite leads to the (known) relation $\theta=1-1/d_w$. This finally leads to the exact relation:
\begin{align}
&S_0=\frac{\overline{T}}{K (1-1/d_w)}\left( \, \int_0^{1} du \,   \int_{1-u}^{\infty} dv \, \frac{G(u/v)}{u^{1/d_w}v^{2-1/d_w}} \right)^{-1}
\label{formuleSFin}.
\end{align}% TRY TO PERFORM ONE INTEGRATION IN THE ABOVE FORMULA
We stress that the dependency of $S_0$ on the source-to-target distance, even when not-trivial, is entirely contained in the term  $\overline{T}$. Indeed, the scaling function $G$  depends only on the large scale properties of the random walk and not on the geometrical parameters. 
 
% THE ABOVE EQUATION MAY BE SKIPPED IF NEEDED, IT IS ALMOST THE SAME AS THE NEXT ONE

While the exact determination of $G$ is a  challenging task, the following decoupling approximation turns out to be very accurate. In this approximation, the return probability to the target at a time $t$ after the first passage time is  independent of the actual value of the FPT, which leads to  $p(0,t+\tau| \text{FPT}=\tau)\simeq p_\pi(t)$ for self-consistence reason. Within this decoupling approximation, $G\simeq 1$ and we obtain 
\begin{equation}
S_0\simeq\frac{\sin(\pi / d_w)}{K \pi} \overline{T} \label{GenEq}, 
\end{equation}
which generalizes Eq.~\eqref{prefMarkT} to non-Markovian processes. 
We now comment on the validity of this key relation. 

%Several comments on the decoupling approximation leading to (\ref{GenEq}) are in order.
First, we stress that Eq.~(\ref{GenEq}) is exact for processes with finite memory time  (i.e. for which the correlation function of increments decays exponentially  at long times). This comes from the very definition of the function $G$, which involves only large time scales in Eq.~(\ref{DefG}), over which this finite memory time becomes irrelevant. This case is illustrated  here by considering a Gaussian process whose Mean Square Displacement function $\psi(t)=\langle [x(t+\tau)-x(\tau)]^2\rangle$ is given by $\psi(t)=Dt+B(1-e^{-\lambda t})$. 
This ``bidiffusive'' process involves two diffusive behaviors at long and short time scales, and presents only one relaxation time $\lambda^{-1}$. This is typically relevant to tracers moving in viscoelastic Maxwell fluids~\cite{grimm2011brownian}, nematics~\cite{turiv2013} or solutions of non-adsorbing polymers~\cite{ochab2011scale}. 
We also consider the effect of multiple relaxation times with the case that $x(t)$ is the position of the first monomer of a flexible polymer chain with $N$ monomers, in the most simple (Rouse, bead-spring) polymer model.  We use recently obtained estimates of $\overline{T}$ in Ref. \cite{guerin2016mean} to obtain estimates of $S_0$ through Eq.~(\ref{GenEq}) and compare with numerical simulations  on Fig.~2(b),(c). We also compare with a pseudo-Markovian approximation  (using Eq.~(\ref{prefMark}) with effective ``propagators'' \cite{PM}). 
Our prediction for $S_0$ is in good agreement with numerical simulations, and shows that even if the memory time is finite, memory effects are strong.

Second, it is showed in SI  that  Eq.~(\ref{GenEq}) is also exact  at first order in $\varepsilon=H-1/2$ for the fractional Brownian Motion (FBM), which is an emblematic example of processes with infinite memory time. 
The FBM is used in fields as varied as hydrology~\cite{mandelbrot1968noah}, finance~\cite{cutland1995stock} and biophysics~\cite{ernst2012fractional,burnecki2012universal}. This Gaussian process is characterized by  $\left\langle [x(t+\tau)-x(t) ]^2\right\rangle = \kappa \tau^{2H}$, with $0<H<1$. 

Third, in the strongly non-Markovian regime, where $\varepsilon$ cannot be considered as small, it turns out that  Eq.~(\ref{GenEq}) provides a very accurate approximation [Fig.~2(e)] of $S_0$, which takes the explicit form 
\begin{align}
S_0=\beta_H  \sin(\pi H)\sqrt{\frac{2}{\pi}}\left(\frac{r_0}{\kappa^{1/2}}\right)^{\frac{1}{H}-1}
\end{align}
where $\beta_H$ is a function of $H$ analyzed in Ref.~\cite{guerin2016mean}. It can indeed be seen on Fig.~2(e) that Eq.~(\ref{GenEq}) correctly predicts the long time behavior of $S(t)$ when $H=0.34$. For this value non-Markovian effects are strong, as can be seen by comparing with the prediction of the pseudo-Markovian approximation, which is wrong by more than one order of magnitude [Fig.~2(e), dashed line]. The value of  $S_0$ is slightly underestimated in the decoupling approximation, but can be made more precise by evaluating the scaling function $G$ (see SI).

Furthermore, our approach also holds in dimension higher than one, even for strongly correlated non-Markovian processes. Indeed, the $d-$dimensional version of Eq.~(\ref{GenEq}) [i.e. Eq.~(\ref{prefMarkT})] correctly predicts (but slightly underestimates) the value of $S_0$ for an example of two dimensional FBM [Fig.~2(f)]. In this example, the target radius $a$ is not zero even if the $a\to 0 $ limit is well defined for compact processes; the dependence of $S_0$ on the target radius is predicted to be the same as that of $\overline{T}$, which is available in the non-Markovian theory of Ref.~\cite{guerin2016mean}.
Finally, in the case of processes with finite memory, we find that Eq.~(\ref{GenEq}) also holds for non-stationary initial conditions. This is illustrated by considering the case of a flexible phantom polymer for which all monomers are placed initially at $r_0$ (instead of having the shape of a random equilibrium coil for stationary initial conditions). This non-stationary initial condition induces  transiently aging dynamics, and $S_0$ is changed with respect to the case of stationary initial conditions, but is still predicted correctly by Eq.~(\ref{GenEq}) [see Fig.~2(d)]. 
 
Finally, let us mention the case of the 1-dimensional run and tumble process, where a particle switches between phases of constant velocities $\pm v$ with  rate $\alpha$. This process is smooth and is a priori not covered by our analysis. However, our relation (\ref{GenEq}) between $S_0$ and $T/V$ is still exact, as is made clear by comparing the results for the mean FPT in confinement  \cite{Masoliver1986} and in semi-infinite space  \cite{Malakar2018}. 
This agreement holds even for non-stationary initial conditions, where the probability $p$ that the initial velocity is positive differs from $1/2$: in this case, one can obtain   $S_0=(r_0+pv/\alpha)\sqrt{2\alpha/(\pi v^2)}=\overline{T}/(K\pi)$, with $K=\sqrt{\alpha/(2\pi v^2)}$, and we can check that our relation still holds \cite{Masoliver1986,Malakar2018}. Furthermore, it also holds in the case of partially reflecting targets, as can be deduced from the results of \cite{Angelani2015}. This suggests that our analysis can be extended to smooth non-Markovian processes with partial absorption as well. 
   
\vspace{1cm}
\noindent  {\bf   Discussion}  

\noindent The determination of  the survival probability $S(t)$, and in particular its dependence on the initial distance to the target,  requires the knowledge of its prefactor $S_0$, which has remained an elusive quantity up to now.   In this article, we have bridged this gap by identifying  a general relation between the long time persistence and the mean FPT in confinement. The latter can be calculated with various recently introduced methods,  for a large class of Markovian 
 \cite{Condamin2007,Benichou2008,Schuss2007} and non-Markovian random walks \cite{guerin2016mean}.  Our theory holds for compact, unbiased walks with stationary increments that are  scale invariant at long times (without confinement), with moments of the position that diverge with time.  Our main result is Eq.~(\ref{prefMarkT}), which is exact for both Markovian processes (such as diffusion in fractals) and for non Markovian processes with finite memory time (for which memory effects are nevertheless quantitatively non-negligible). For long-ranged correlated processes such as FBM our formula provides a good  approximation of $S_0$ in one or higher dimensions, and is found to be exact at first order in a perturbation expansion around Brownian motion. 
Together, our results thus improve our understanding of the impact of memory on the statistics of long first passage events.

 \vspace{0.5cm}
{\bf Acknowledgments.} This work was supported by ERC grant FPTOpt-277998. Computer time for this study was provided by the computing facilities MCIA (Mesocentre de Calcul Intensif Aquitain) of the Universit\'e de Bordeaux and of the Universit\'e de Pau et des Pays de l'Adour.
   
\vspace{0.5cm}
{\bf Data and code availability.}   
  The numerical data presented in Figure 2, as well as the code that generated these data,  are available from the corresponding author on reasonable request.

  \vspace{0.5cm}
{\bf Competing interests statement.} 
  The Authors declare no competing interests.
   
%\bibliography{BiblioPostDoc}

\begin{thebibliography}{10}
\bibitem{PM}
\bibinfo{howpublished}{The pseudo-Markovian approximation, which is similar to
  the Wilemski-Fixman approximation for the polymer cyclization kinetics
  problem, consists in using effective propagators in Eq.~(\ref{GenEq}), i.e
  $p(\ve[x],t\vert
  \ve[x]_0)=e^{-(\ve[x]-\ve[x]_0)^2/2\psi(t)}/(2\pi\psi(t))^{d/2}$.}
\end{thebibliography}
%\bibliographystyle{naturemag}

\newpage

\begin{figure}[h!]
 \includegraphics[width=10cm]{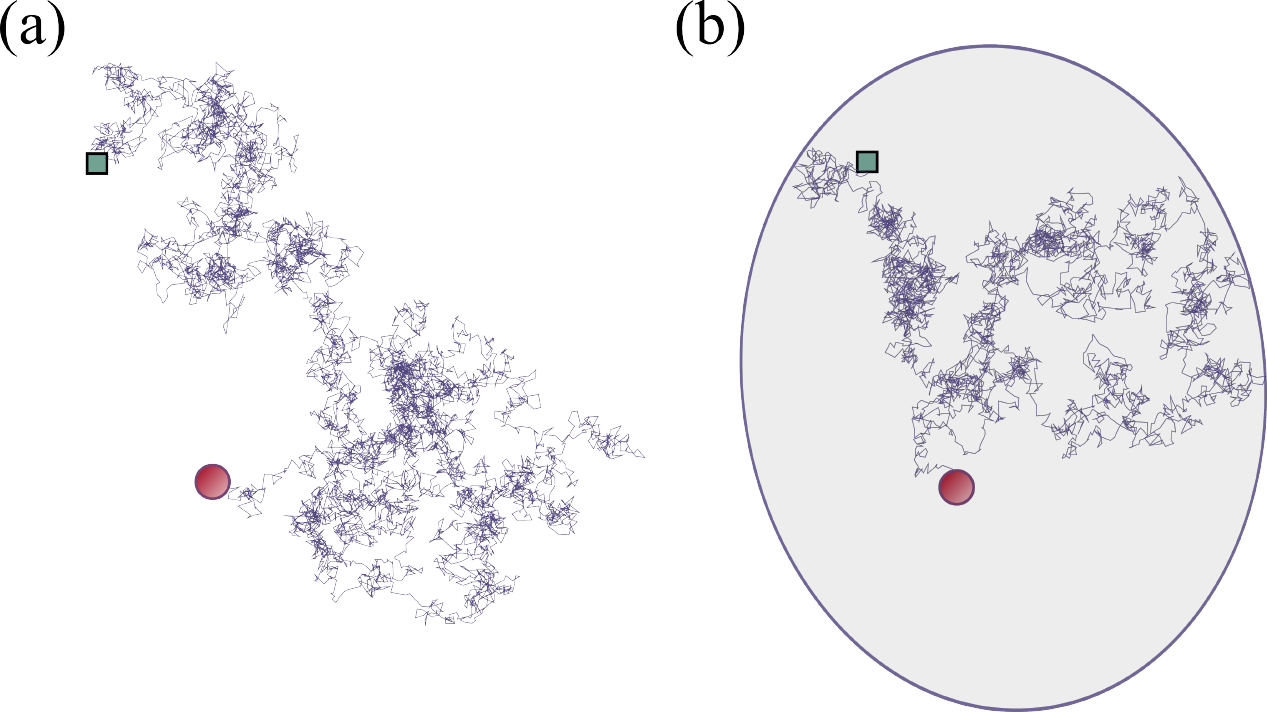} 
\caption{\textbf{First-passage problem with or without confinement}. Two first passage problems in which a random walker starting from a given site (green square) reaches a target (red disk) at the end of a stochastic trajectory: (a)  in free space, (b)  in a confined reflecting domain. Sample trajectories for fractional Brownian motion ($H=0.45$) are shown. }
\label{PrefFig1}
\end{figure}

\newpage

\begin{figure}[!h] 
 \includegraphics[width=10cm]{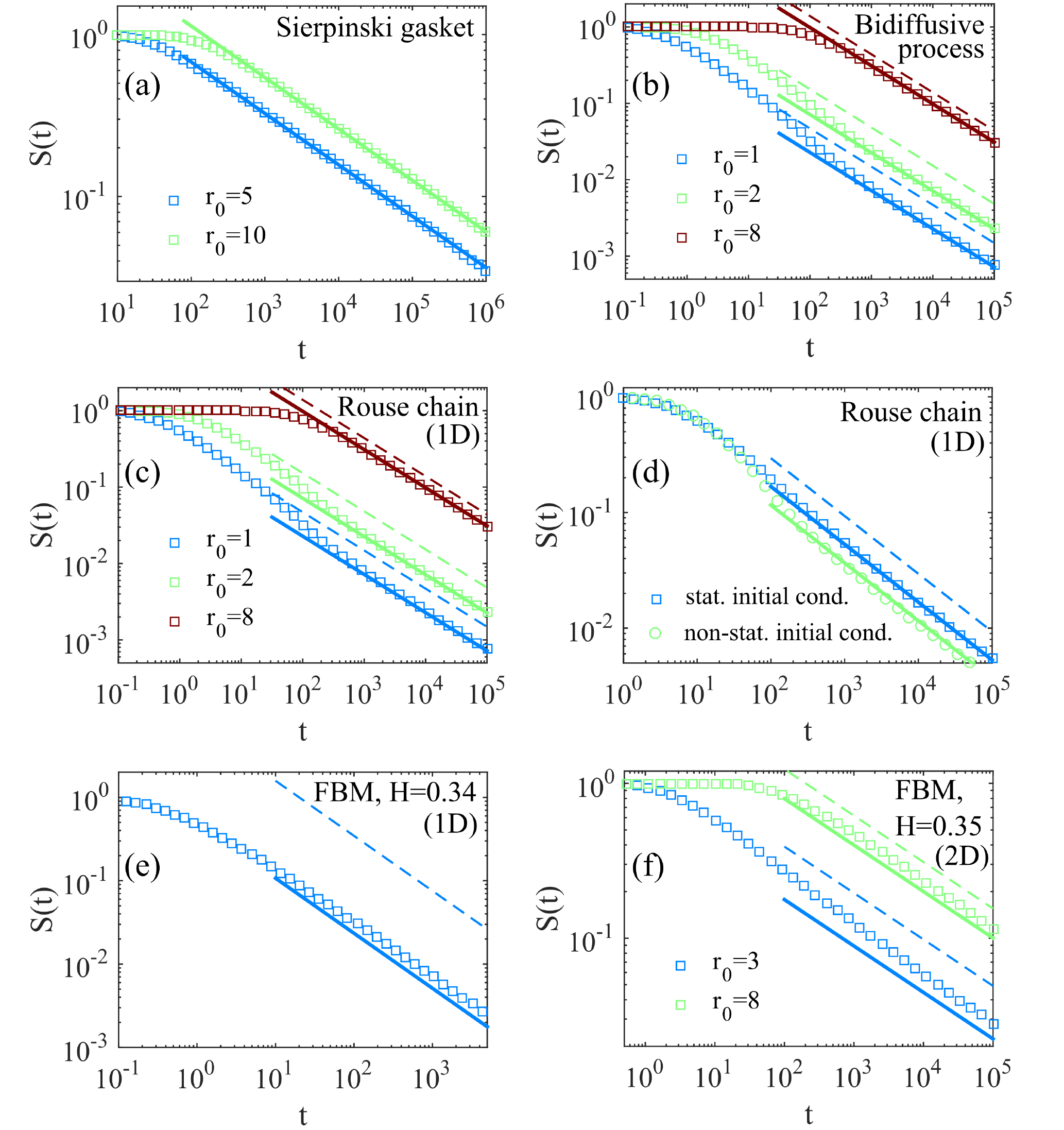} 
 \caption{\textbf{Survival probability $S(t)$ for various stochastic processes}. In all graphs, symbols are the results of stochastic simulations (detailed in SI), continuous lines give the  theoretical predictions  [Eq.~(\ref{GenEq})], and dashed line represent the predictions of the pseudo-Markovian approximation~\cite{PM}. (a) $S(t)$ for a random walk on the Sierpinski gasket for two values of the initial (chemical) source-target distance. Here, $d_f=\ln3/\ln2$, $d_w=\ln5/\ln2$, and  $K\simeq 0.30$   \cite{weber2010random,meroz2011distribution}. Simulations are shown for a fractal of generation $11$. Continuous lines are the predictions of Eq.~(\ref{ZeroConstantS0}). (b) $S(t)$ for a 1-dimensional  ``bidiffusive'' Gaussian process of MSD $\psi(t)=t+30(1-e^{-t})$. (c) $S(t)$ for a one-dimensional Rouse chain with $N=20$ monomers, for various source-to-target distance $r_0$ (indicated in the legend in units of the monomer length). (d) $S(t)$ for the same system with $N=15$ and $r_0=3$, comparing stationary initial conditions (the other monomers being initially at equilibrium) or non-stationary ones (for which all monomers start at the same position $r_0$). (e) $S(t)$ for a one dimensional FBM of MSD $\psi(t)=t^{2H}$ with Hurst exponent $H=0.34$. (f) Two-dimensional FBM of MSD $\psi(t)=t^{2H}$ in each spatial direction with $H=0.35$. The target is a disk of radius $a=1$ and $r_0$ is the distance to the target center. For (b),(c),(d),(e),(f) the continuous lines represent our predictions [Eq.~(\ref{GenEq})], in which $\overline{T}$ is calculated by using the theories of Refs.~\cite{Guerin2012a,guerin2016mean,Guerin2013}; in (b) and (c) the only hypothesis to predict $\overline{T}$ is that the distribution of supplementary degrees of freedoms at the FPT is Gaussian, in (e) and (f) we use the additional  ``stationary covariance'' approximation. In (d), for non-stationary initial conditions, $\overline{T}$ is measured in simulations in confined space. A table that compares the values of $S_0$ in the theory and in the simulations  is given in SI.}
\label{PrefFig2}
\end{figure}

\end{document}